\documentclass[conference]{IEEEtran}
\IEEEoverridecommandlockouts

\usepackage{amsmath,amssymb,amsfonts}
\usepackage{algorithmic}
\usepackage{algorithm}
\usepackage{graphicx}
\usepackage{textcomp}
\usepackage{xcolor}
\usepackage{xspace}
\usepackage{scalefnt}
\usepackage{physics}
\usepackage[hidelinks]{hyperref}
\usepackage{tabularx}
\usepackage{booktabs}

\usepackage[noadjust]{cite}
\newcommand\smaller[2][0.85]{{\scalefont{#1}#2}}

\newcommand{\FRONTIERE}{\mbox{Frontier-E}\xspace}
\newcommand{\CRKHACC}{\smaller{CRK-HACC}\xspace}

\newcommand{\CRKSPH}{CRKSPH\xspace}

\renewcommand{\thefootnote}{\fnsymbol{footnote}}

\newcommand{\Frontier}{Frontier\xspace}
\newcommand{\Orion}{Orion\xspace}
\newcommand{\Aurora}{Aurora\xspace}
\newcommand{\AMD}{AMD\xspace}
\newcommand{\Intel}{Intel\xspace}
\newcommand{\Nvidia}{Nvidia\xspace}
\newcommand{\MIX}{MI250X\xspace}
\newcommand{\TFLOPS}{TFLOPs\xspace}
\newcommand{\EFLOPS}{EFLOPs\xspace}

\renewcommand{\thefootnote}{\fnsymbol{footnote}}

\begin{document}
\bstctlcite{IEEEexample:BSTcontrol}

\title{Cosmological Hydrodynamics at Exascale: \\ A Trillion-Particle Leap in Capability}

\author{
\IEEEauthorblockN{
Nicholas~Frontiere\IEEEauthorrefmark{1}, 
J.D.~Emberson\IEEEauthorrefmark{1}, 
Michael~Buehlmann\IEEEauthorrefmark{1}, 
Esteban~M.~Rangel\IEEEauthorrefmark{1}, \\
Salman~Habib\IEEEauthorrefmark{1}\IEEEauthorrefmark{2},
Katrin~Heitmann\IEEEauthorrefmark{2},
Patricia~Larsen\IEEEauthorrefmark{1},
Vitali~Morozov\IEEEauthorrefmark{3},
Adrian~Pope\IEEEauthorrefmark{1}, \\
Claude-Andr\'e Faucher-Gigu\`ere\IEEEauthorrefmark{4},
Antigoni~Georgiadou\IEEEauthorrefmark{5},
Damien~Lebrun-Grandi\'{e}\IEEEauthorrefmark{6},
Andrey~Prokopenko\IEEEauthorrefmark{6}
}
\IEEEauthorblockA{\IEEEauthorrefmark{1}Computational Science Division, Argonne National Laboratory}
\IEEEauthorblockA{\IEEEauthorrefmark{2}High Energy Physics Division, Argonne National Laboratory}
\IEEEauthorblockA{\IEEEauthorrefmark{3}Argonne Leadership Computing Facility, Argonne National Laboratory}
\IEEEauthorblockA{\IEEEauthorrefmark{4}Department of Physics and Astronomy, Northwestern University}
\IEEEauthorblockA{\IEEEauthorrefmark{5}National Center for Computational Sciences, Oak Ridge National Laboratory}
\IEEEauthorblockA{\IEEEauthorrefmark{6}Computational Sciences and Engineering Division, Oak Ridge National Laboratory}
\IEEEauthorblockA{Emails: \{nfrontiere, jemberson, mbuehlmann, erangel\}@anl.gov, \\
                          \{habib, heitmann, prlarsen, morozov, apope\}@anl.gov\\
                    cgiguere@northwestern.edu, \{georgiadoua, lebrungrandt, prokopenkoav\}@ornl.gov}

}

\maketitle
\vspace{-0.25cm}
\begin{abstract}
Resolving the most fundamental questions in cosmology requires simulations that match the scale, fidelity, and physical complexity demanded by next-generation sky surveys. To achieve the realism needed for this critical scientific partnership, detailed gas dynamics, along with a host of astrophysical effects, must be treated self-consistently with gravity for end-to-end modeling of structure formation. As an important step on this roadmap, exascale computing enables simulations that span survey-scale volumes while incorporating key subgrid processes that shape complex cosmic structures. We present results from \CRKHACC, a cosmological hydrodynamics code built for the extreme scalability requirements set by modern cosmological surveys. Using separation-of-scale techniques, GPU-resident tree solvers, in~situ analysis pipelines, and multi-tiered I/O, \CRKHACC executed \FRONTIERE: a four trillion particle full-sky~simulation, over an order of magnitude larger than previous efforts. The run achieved 513.1\,PFLOPs peak performance, processing 46.6 billion particles per second and writing more than 100\,PB of data in just over one week of runtime. 
\end{abstract}

\section{Simulation Introduction}\label{sec:justification}

The four trillion particle \FRONTIERE simulation was carried out with the GPU-accelerated cosmological hydrodynamics \CRKHACC code using a 4.7~Gpc simulation box, with an equal number of baryonic and dark matter tracer particles. It achieved 513\,PFLOPs peak performance on 9{,}000 nodes of Oak Ridge National Laboratory's \Frontier system, processing 46.6 billion simulation particles/second. The run generated~$\mathord{>}$100\,PB of data in under 3\% of the total runtime, establishing a new standard of end-to-end performance for large-scale, multi-physics cosmological simulations -- including compute, I/O, and scalability. Predictions for cosmological observables and probes, many computed {\em in situ},  cover a wide range of wave bands, from radio to X-ray.\\

\section{Overview of the Problem}\label{sec:overview}

Understanding and predicting the formation and evolution of structure in the Universe is a central theme of modern cosmology. Seeded by tiny density perturbations imprinted at the earliest epochs, gravitational collapse in the expanding Universe gives rise to a vast distribution of dark matter and ionized gas that surrounds galaxies and galaxy clusters, forming an intricate network of filaments and nodes known as the cosmic web.
Among the most important open questions in the field are the origin of primordial fluctuations, the nature and role of dark matter (the dominant mass component of the Universe), the cause of late-time cosmic acceleration (e.g., dark energy or modified gravity), and how ordinary (baryonic) gas interacts with these elements to shape the Universe we observe today.

\begingroup
\renewcommand{\thefootnote}{\arabic{footnote}}
\setcounter{footnote}{0}
A new generation of high-precision cosmological surveys, such as the Dark Energy Spectroscopic Instrument\footnote{\url{https://www.desi.lbl.gov}}, Euclid\footnote{\url{https://www.euclid-ec.org}}, the Roman Space Telescope\footnote{\url{https://roman.gsfc.nasa.gov}}, the Vera C. Rubin Observatory\footnote{\url{https://www.lsst.org}}, and SPHEREx\footnote{\url{https://spherex.caltech.edu}}, are designed to measure and characterize the change in matter distribution over time. Detailed cosmological simulations are necessary for interpreting results from these observations and making predictions for models that go beyond current theoretical assumptions. This need is especially pressing as the standard cosmological model -- $\Lambda$CDM -- faces increasing tension across multiple observables (e.g., Ref.~\cite{efstathiou2025challenges}), requiring simulations capable of disentangling potential new physics from systematic effects and baryonic contributions.

\textcolor{red}{}

\endgroup
\setcounter{footnote}{0}
Until now, survey-scale simulations have been limited to modeling the evolution of structure using gravity-only N-body  approaches with trillions of particle tracers in gigaparsec-scale volumes -- equivalent to billions of light-years across~(e.g., Refs.~\cite{potter2017pkdgrav3,heitmann2019}). Although these simulations provide valuable insights, they neglect gas dynamics and astrophysical feedback processes, both of which produce signals that modern observations are increasingly sensitive to; understanding these processes is a key issue in improving the sensitivity, accuracy, and robustness of several cosmic probes.

To improve simulation fidelity, cosmological hydrodynamic simulations that evolve both gas and dark matter using accurate fluid dynamics solvers are widely used~\cite{vogelsberger2020}, but are at least 10 to 20 times more computationally expensive than gravity-only runs. Consequently, performing high-resolution, full-sky hydrodynamic simulations has remained out of reach -- not only due to the extreme computational demands, but also because few cosmology codes are capable of both scaling efficiently on leadership-class high-performance computing (HPC) systems and of effectively exploiting GPU hardware.

The advent of exascale machines has introduced the computational capability that was previously missing to carry out larger-scale hydrodynamic simulations with significantly reduced runtimes. With more than an order-of-magnitude increase in parallel throughput, these systems make it possible -- at least in principle -- to perform state-of-the-art simulations at the same scale as their gravity-only predecessors with realistic wall-clock times (roughly days to weeks of machine time).

We present the results of the Frontier Exascale simulation (\FRONTIERE), the first exascale cosmological run of its kind. Executed on the \Frontier supercomputer, \FRONTIERE evolves \emph{four trillion particles}, evenly split between baryonic gas and dark matter, within a cubic simulation volume exceeding 100\,$\mathrm{Gpc}^3$, or about 15.3 billion light-years on a side. Built using the \CRKHACC framework~\cite{habib2013hacc,frontiere2022simulating}, the simulation employs a separation-of-scale gravity solver, a higher-order particle-based hydrodynamic implementation, and incorporates detailed astrophysical source models. Specifically, \CRKHACC includes treatment of radiative and metal-line cooling, star formation and supernova feedback, stellar chemical enrichment, and active galactic nuclei (AGN) feedback; these models require much finer temporal resolution, induce the formation of stars and galaxies, and inject large amounts of energy in the simulation. 

The impact of the results from \FRONTIERE is remarkably broad: the simulated volume is large enough to provide statistically converged measurements for all clustering probes, the simulation spans the full redshift range of cosmic history targeted by all major large-area surveys, and the included physics enables realistic predictions for observables across the X-ray, optical, infrared, mm-wave, and radio bands. In fact, \FRONTIERE was conceived within the Exascale Computing Project\footnote{\url{https://www.exascaleproject.org/}} (ECP) as a “grand challenge problem” of scientific inquiry -- designed to demonstrate impactful research achievable only on exascale machines. It is one of the first such efforts to successfully complete on a realized exascale system. 

A key advantage of \FRONTIERE is the ability to make joint predictions across cosmological probes -- a critical test of the consistency of the physical modeling. The number of cosmological objects in \FRONTIERE is also unprecedented; for example, it contains roughly 570{,}000 galaxy clusters, compared to fewer than 50{,}000 currently observed. This makes it possible to study not only the mean properties of these structures, but also their full distribution in detail.

Fully utilizing an exascale machine for cosmological predictions at the scale of \FRONTIERE presents several major challenges: (1) scalability; (2) significant I/O demands; (3) realistic time-to-solution; (4) fault tolerance; and (5) performance and portability. In this paper, we describe how each of these challenges was directly addressed in the development of \CRKHACC, with \FRONTIERE serving as a demonstration of the capabilities and scientific impact that exascale systems deliver when pushed to their limits. The methods developed, including GPU-resident tree solvers, optimized interaction kernels, and multi-tiered I/O, are generalizable to fields that use particle interaction solvers -- such as beam dynamics, plasma physics, and molecular dynamics, which can apply similar strategies to achieve high performance and throughput.

\section{Current State of the Art}\label{sec:SOTA}

As outlined in a tri-agency (DOE, NASA, NSF) report on the cosmological simulation landscape~\cite{battaglia2020}, upcoming survey predictions require gigaparsec-scale box volumes to achieve the statistical precision needed for comparison with observations, along with sufficiently high mass resolution to resolve the faintest objects of interest. Together, these demands translate into simulations that evolve \emph{trillions} of particles.

Gravity-only N-body simulations have successfully exploited HPC machines to meet the extreme volume and resolution requirements for survey-scale predictions, reaching the particle counts dictated by these considerations. 
These simulations simultaneously sample the largest cosmic structures and resolve compact, collapsed dark matter-dominated clumps known as halos.
Halos form hierarchically, with smaller structures merging to form larger ones. Galaxies form within these halos: the most massive can host hundreds to thousands of galaxies, while smaller halos typically contain the galaxies that dominate survey observations.
N-body simulations are widely used to generate synthetic observables, including mock sky maps and galaxy catalogs, which play a central role in the analysis pipelines of modern surveys~\cite{korytov2019,castander2024euclid}.

Hydrodynamic simulations, on the other hand, have made tremendous progress in capturing the complex interplay of baryons and dark matter, including gas cooling, star formation, and AGN feedback (see, e.g. Ref.~\cite{vogelsberger2020} for a review). However, no results to date have achieved the combination of volume and resolution required to match their gravity-only counterparts at the scale necessary for large-scale optical surveys.

\begin{figure}
    \includegraphics[width=\linewidth]{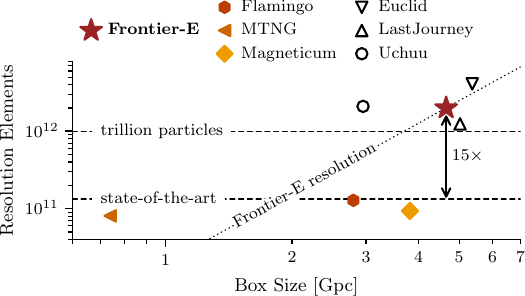}
    \caption{Comparison of large-volume simulations for gravity-only (black markers) and state-of-the-art cosmological hydrodynamics solvers (colored markers). The \FRONTIERE simulation is the first to break the trillion-particle barrier, reaching the same scale as leading gravity-only counterparts. \emph{Resolution Elements} refers to the count of dark matter–baryon particle pairs in hydrodynamic simulations, to allow fair comparison with single-species gravity-only runs. The dotted line indicates the particle count required to match the mass resolution of \FRONTIERE as a function of simulation volume.  
}
    \label{fig:frontiere_in_context}
\end{figure}

In Figure~\ref{fig:frontiere_in_context}, we compare several state-of-the-art cosmological hydrodynamic simulations: FLAMINGO~\cite{schaye2023flamingo}, MillenniumTNG~\cite{pakmor2023}, and Magneticum~\cite{dolag2016sz}. For reference, we also include modern gravity-only campaigns from the Euclid Flagship simulation run with \smaller{PKDGRAV3}~\cite{potter2017pkdgrav3}, the Last Journey simulation~\cite{heitmann2021}, and the Uchuu simulation~\cite{ishiyama2021uchuu}. The x-axis denotes the comoving simulation box length in gigaparsecs (Gpc), while the y-axis shows the total number of resolution elements, defined as dark matter–baryon particle pairs in hydrodynamic simulations to ensure consistency with single-species gravity-only runs.\footnote{Typical modern hydrodynamic simulations, including \FRONTIERE, represent baryons and dark matter with equal numbers of particles. Aside from the additional computational complexity, such runs require at least twice the memory of gravity-only simulations.}

The largest of the previous hydrodynamic simulations have only just surpassed the hundred-billion baryon particle mark -- still more than an order of magnitude below the scale required for full-survey fidelity. Moreover, none of the previously reported large-scale hydrodynamic simulations utilize GPU-accelerated solvers, a limitation in an era increasingly defined by GPU-based high-end HPC systems.

The \FRONTIERE simulation represents a leap forward in capability. It is the first exascale-class hydrodynamic simulation, evolving a total of four trillion particles -- more than a 15-fold increase over previous efforts -- and achieving higher resolution than the two largest-volume hydrodynamic simulations to date. As seen in Figure~\ref{fig:frontiere_in_context}, \FRONTIERE reaches the predictive scales previously attained only by gravity-only simulations. Its immense volume is essential for embedding synthetic observations within a single, self-consistent domain and for generating statistically meaningful, full-sky, multi-wavelength predictions. 

\FRONTIERE is the first large-scale hydrodynamic simulation to date that both harnesses GPUs and scales efficiently to a full exascale-class system. Achieving this required the convergence of several critical capabilities: algorithmic advances; sufficient system memory to evolve trillions of particles; a robust I/O subsystem to support writing over 12\,PB of scientific output along with continuous checkpointing ($\mathord{>}90$\,PB) for fault tolerance, particularly important given the high interruption rates of modern machines~\cite{kokolis2024}; and the compute power necessary to reach a feasible time-to-solution -- on the order of one week of machine time.  This combination of system-scale resources and extensive software development marks a new era in survey-scale cosmological hydrodynamics, only made possible by exascale platforms.

\section{Innovations Realized}\label{sec:innovation}

\begin{figure*}[t]

    \includegraphics[width=\linewidth]{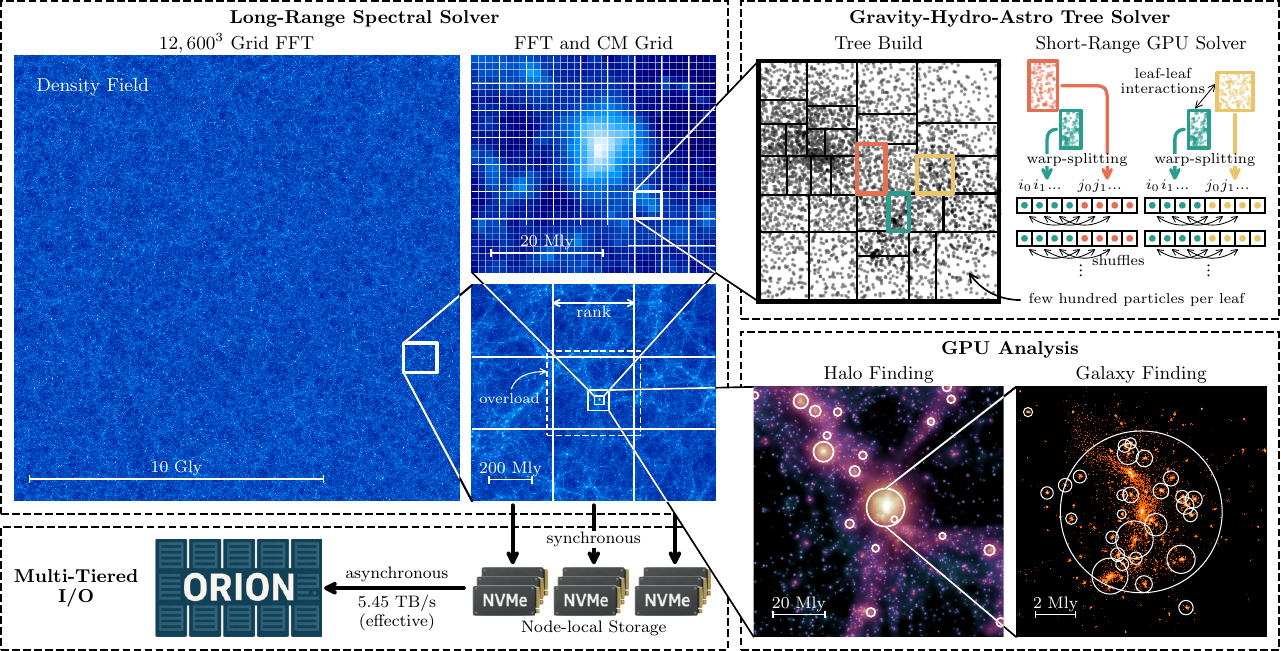}
    \caption{\CRKHACC architecture diagram of the primary simulation components, spanning gigalight-year volumes down to short-range forces acting on individual particles. The distributed long-range spectral FFT solver operates over the global domain across all nodes (top left). After k-d trees are constructed in chaining mesh bins, the entire overloaded rank is pushed to the GPU (top right), where short-range force operators process leaf-leaf interactions using warp-splitting kernels. Cluster-based in situ analysis is also GPU-accelerated (bottom right). Multi-tier I/O (bottom left) outputs data using synchronous writes to node-local NVMe SSDs, which bleed data to the PFS asynchronously. For \FRONTIERE, the time-to-solution contributions from the long-range solver, tree build, short-range solver, in situ analysis, and I/O were \{1.7\%, 1.7\%, 79.6\%, 11.6\%, and 2.6\%\}, respectively. Over 90\% of solver time was executed on the GPU; see Section~\ref{subsec:TTS}.}
    \label{fig:schematic}
\end{figure*}

The \CRKHACC framework incorporates several innovations necessary to fully exploit exascale systems, particularly to execute the \FRONTIERE simulation. We begin with an overview of the code's architecture -- a multiscale, hybrid solver designed for high performance on modern HPC platforms. Special emphasis is placed on the GPU-resident implementation of short-range operators, which facilitates high performance on accelerated hardware. We then highlight several key innovations that directly address the major challenges of survey-scale cosmological hydrodynamic simulations: I/O, time-to-solution, and sustained performance. For further details not provided in this synopsis, see Refs.~\cite{habib2013hacc,frontiere2022simulating}. 

As highlighted below, the techniques described are generalizable to Lagrangian-based codes (e.g., particle-in-cell methods in plasma physics, molecular dynamics with pairwise force kernels, etc.) and are not exclusive to cosmology. The \CRKHACC solver was designed with HPC bottlenecks and architectural challenges in mind, rather than as a custom solution for a single scientific application.

\subsection{Architecture Overview}\label{subsec:arch}

An overview of the \CRKHACC framework is shown in Figure~\ref{fig:schematic}. The full 15.3\,Gly simulation volume is divided into cuboid subdomains, each evolved independently on individual compute ranks. These regions overlap at their boundaries, where particles are duplicated (``overloaded'') to enable short-range force computations to remain node-local, eliminating the need for MPI communication -- similar in spirit to ghost zones in mesh-based solvers.

The intermediate and long-range gravitational interaction is computed using a spectral (FFT-based) solver for the Poisson equation via the particle-mesh (PM) approach~\cite{hockney1988}. To achieve the required accuracy directly would demand grids with millions of cells per dimension -- far beyond the capacity of current supercomputers. The gravitational field is therefore decomposed into long- and short-range components, where the short-range forces are evaluated locally using direct or approximate (tree-based) particle methods. CRK-HACC uses a specially designed, high-order spectrally-filtered PM method enabled by a high-performance distributed FFT implementation, called SWFFT.\footnote{We have made SWFFT publicly available at \url{https://git.cels.anl.gov/hacc/SWFFT}} This approach allows low-noise handover to the short-range solver on a compact spatial scale, further improving the total solver performance~\cite{habib2013hacc}.

The PM solver operates on a global mesh of size \mbox{$N=12{,}600^3$} for the \FRONTIERE run, corresponding to two trillion cells distributed across all nodes (top left panel of Figure~\ref{fig:schematic}). Once the global gravitational field is computed, overloaded rank domains are transferred to the GPU, where short-range forces (including hydrodynamics and astrophysical feedback) are evaluated. This approach minimizes communication costs and leverages GPU acceleration for all local interactions. Further, the multi-scale design supports mixed precision, where the FFT-based long-range solver runs in FP64 to preserve spectral accuracy, while the short-range GPU solver can be executed in FP32, gaining performance and memory efficiency without compromising scientific fidelity.

For baryonic gas dynamics, \CRKHACC uses a mesh-free, higher-order smoothed particle hydrodynamics (SPH) method known as Conservative Reproducing Kernel SPH (\CRKSPH)~\cite{frontiere2017}. This formulation solves the inviscid Euler equations using particle-based interpolants. \CRKSPH explicitly conserves mass, momentum, and energy, while reducing numerical diffusion and accurately modeling shocks and fluid instabilities. The solver has been shown to produce consistent results compared to adaptive mesh refinement (AMR) codes in cosmological fluid simulations~\cite{frontiere2022simulating,chabanier2022modeling}.

Beyond gas dynamics, the simulation includes subgrid astrophysical models for star formation, metal enrichment, and feedback from supernovae and AGN, calibrated to observations.\footnote{For \FRONTIERE, these models were calibrated using a suite of mid-scale simulations run on Perlmutter at NERSC.} These modules, while computationally expensive, are necessary for resolving high-density regions and introducing rapid gas collapse. The timescales involved are much shorter than those for gravitational evolution, requiring smaller integration steps and increasing computational cost.

To resolve these local processes without reducing the global timestep, we employ an adaptive integration scheme~\cite{saitoh2010}. Particles are grouped into timestep bins and evolved according to local conditions, resulting in heterogeneous workloads across the domain. This increases the depth of the integration loop relative to fixed-timestep, gravity-only simulations and is efficiently handled on the GPU, as described in Section~\ref{subsec:warp_split}.

The \FRONTIERE simulation evolves through 625 global PM timesteps, during which each rank locally integrates all short-range interactions -- including hydrodynamics, subgrid physics, and feedback -- capturing the full history of structure formation in both baryonic gas and dark matter. These local integrations can involve thousands of subcycled time steps per PM interval, reflecting the fine-grained time resolution required to model astrophysical processes.

The \CRKHACC solver includes roughly fifty computational kernels that implement short-range operators, including astrophysical feedback modules. These were developed for accelerated execution using a GPU-resident approach in which data remains on the device throughout each PM timestep, minimizing transfers to and from the host. The ten most compute-intensive functions -- particularly those responsible for hydrodynamics and gravitational forces -- have been heavily optimized and make use of the warp splitting technique discussed in Section~\ref{subsec:warp_split}.

All GPU kernels use custom abstracted function call interfaces that map to vendor-specific languages (CUDA, HIP, and SYCL), enabling GPU portability. Performance across hardware vendors is studied in detail in Ref.~\cite{rangel2023}, while Section~\ref{subsec:perf_port} demonstrates sustained performance of the \FRONTIERE workload on \Intel, \AMD, and \Nvidia GPUs.

\subsection{Key Innovations}\label{subsec:key_innov}

Given the overview of the \CRKHACC framework, we highlight four important innovations that are responsible for addressing the significant performance, scientific analysis, and I/O requirements for a complete end-to-end simulation. These include an optimized tree-based data structure, a customized leaf-interaction splitting approach, an extensive GPU-accelerated in~situ analysis pipeline, and a multi-tier I/O capability.
We again emphasize that all innovations below are generalizable to particle-based approaches and are designed to avoid general system-level bottlenecks, such as complex memory hierarchies, host–device data transfers, kernel performance limitations, and parallel file system (PFS) overheads. 
\vspace{0.2cm}
\subsubsection{GPU Tree Solver}\label{subsec:GPU_tree}

In SPH, fluid quantities are estimated at particle positions via kernel-weighted interpolation over local neighborhoods. In the high-order \CRKSPH formulation, this involves approximately 270 nearby particles per evaluation, requiring efficient spatial search for both performance and scalability. \CRKHACC employs k-d tree spatial decompositions to organize particles and generate interaction lists, with pairwise leaf-to-leaf kernel operations used to evaluate hydrodynamic forces.

Unlike mesh-based codes with fixed computational stencils, the topology of particle interactions in SPH evolves dynamically. As particles move, the tree must be updated to reflect new local neighborhoods. Although this adaptivity is one of the strengths of SPH for resolving structure formation, it presents challenges for GPU execution, where memory access patterns and control flow must be highly structured.

To manage this complexity, the spatial domain of each rank is divided into fixed-size subvolumes called chaining mesh (CM) bins. All short-range forces operate only within a bin and its neighbors. The CM grid is approximately four FFT grid cells wide, as shown in Figure~\ref{fig:schematic}. Each CM bin contains a local k-d tree that subdivides its particles into base-level leaves of a few hundred particles each -- a relatively coarse depth compared to CPU trees built to the single-particle level.

Rather than constructing and storing full hierarchical tree structures, we retain only the base leaves and allow their bounding boxes to grow over time, avoiding dynamic rebuilding. Thus, the chaining mesh and trees are built once per global PM step, and the leaves expand as needed during the short-range evolution. This avoids costly repartitioning, at the expense of increased neighbor overlap.

Combined with the adaptive (hierarchical) timestepping discussed earlier, this design is well suited for GPU acceleration. Only ``active'' leaves are updated at each hydrodynamic substep, and updating bounding boxes and interaction lists is significantly faster than executing the force kernels. As shown in Section~\ref{sec:results}, this enables sustained high performance with the vast majority of runtime spent in compute-dominated force kernels rather than memory-bound tree assembly.
\vspace{0.2cm}

\subsubsection{Warp Splitting}\label{subsec:warp_split}

The primary computational component of the GPU solver is the set of leaf-to-leaf interaction kernels. In these short-range operators, all particles \( i \) in one leaf interact with all particles \( j \) in a neighboring leaf, with both sets typically updated. In more complex physics modules, these kernels are often constrained by register pressure due to the need to store numerous state variables for both particles \( i \) and \( j \) within a single thread.

Most interaction kernels accumulate a pairwise quantity $\phi_i$, generally of the form:
\begin{equation}\label{eqn:arb}
    \phi_i = \sum_j\phi_{ij} = \sum_j f(\alpha_i, \beta_i, \dots, \alpha_j, \beta_j, \dots),
\end{equation}
where \( f \) is a kernel-specific function evaluated across all neighbors \( j \), using contributions of potentially many state variables ($\alpha,\beta, \dots$) from each particle pair. 

Fortunately, the kernel function often contains separable terms, e.g.,  
\begin{equation}\label{eqn:separable}
\phi_{ij} = f_i(\alpha_i, \dots) \ast g_j(\alpha_j, \dots)\ast h_{ij}(|\mathbf{r}_i - \mathbf{r}_j |,\alpha_i, \dots)\cdots,
\end{equation}
where $\ast$ denotes a general arithmetic operation, and $f,g$ and $h$ are components that depend solely on \( i \), solely on \( j \), or on a limited number of coupled variables such as the separation distance, $|\mathbf{r}_i - \mathbf{r}_j|$. This structure is typical for (anti)symmetric kernels, where $\phi_{ij} = \pm \phi_{ji}$, such as the SPH hydrodynamic or gravitational force calculations in \CRKHACC. Importantly, the shared partial terms are redundant when individually computed for both particle \( i \) and particle \( j\); avoiding this duplication reduces register requirements.

To exploit this structure, we introduce a technique called \emph{warp splitting}\footnote{We follow the \Nvidia nomenclature, where a warp is a group of threads that execute the same instruction simultaneously: 32 threads on \Nvidia and \Intel GPUs, and 64 threads on \AMD.}, outlined in Algorithm \ref{algo:warp} and partially visualized in Figure~\ref{fig:schematic} (top right panel). For each interacting leaf pair, a warp is split such that half of its threads represent particles from leaf \( i \), and the other half from leaf \( j \). After two coalesced global memory reads of the relevant particle states, threads repeatedly communicate with their partners using warp shuffles (fast register-level exchanges between threads in the same warp), avoiding potentially expensive memory operations. For example, in Equation~\ref{eqn:separable}, $f_i$ and $g_j$ can be computed independently and exchanged between threads to evaluate $\phi_{ij}$.

\begin{algorithm}
\caption{Warp Splitting Example}
\begin{algorithmic}[1]\label{algo:warp}
\STATE Warp loads particle data from global memory: $\mathbf{r}, \alpha, \dots$ 
       \\(half threads load from leaf $i$, other half from leaf $j$)
\FOR{each partner $j$ in half-warp}
    \STATE Exchange data via warp shuffle: $\mathbf{r}_j$ = shuffle($\mathbf{r}_i$)
    \STATE Eval $f_i(\alpha_i,\dots)$, $g_i(\alpha_i,\dots)$, $h_{ij}(|\mathbf{r}_i - \mathbf{r}_j |,\dots) \cdots$
    \STATE Exchange partials: $g_j$ = shuffle($g_i$)\\
    \STATE Eval $\phi_{ij} = f_i\ast g_j \ast h_{ij} \cdots$ 
    \STATE Accumulate $\phi_i \mathrel{+}= \phi_{ij} $
\ENDFOR
\STATE Perform atomic update of $\phi_i$ to global memory \\
\end{algorithmic}
\vspace{0.3cm}
\begin{tabularx}{\linewidth}{@{}Xr@{}}
    \begin{minipage}[t]{\linewidth}\raggedright Illustration of the first two iterations of warp-shuffle operations on a single split warp\end{minipage}
&
\raisebox{-0.85\totalheight}{\includegraphics[width=0.65\linewidth]{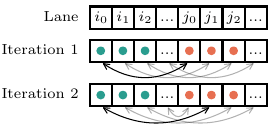}}
\end{tabularx}
\vspace{0.1cm}
\end{algorithm}

Since each thread only stores local state and shares minimal intermediate values (e.g., scalar coefficients or gradients), register pressure is greatly reduced. Each thread will iterate and interact with all threads in the opposite half-warp; thread \( i \) will be assigned a unique partner \( j \) for each iteration. After all unique combinations of particle pairs are evaluated, final results are accumulated locally and written to global memory using leaf-level atomics, minimizing contention.

The warp splitting approach has several performance advantages: (1) register usage is reduced through shared partial computations, (2) expensive global memory access is minimized and coalesced, (3) shuffles enable efficient intra-warp communication, (4) global atomics are localized to per-leaf reductions, and (5) the method generalizes to all \CRKHACC interaction kernels, as well as other particle-based methods with separable or symmetric interaction structures. These include examples from molecular dynamics (e.g., pairwise interactions such as Lennard-Jones or Coulomb potentials~\cite{frenkel2023understanding}) and plasma physics (e.g., collective or screened particle interactions~\cite{chen2015introduction}). Warp splitting is a key optimization that contributes to the high GPU utilization and fast solver time-to-solution observed in Sections~\ref{subsec:TTS} and~\ref{subsec:perf_port}.
\vspace{0.2cm}
\subsubsection{In~situ GPU-Accelerated End-to-End Analysis}\label{subsec:cosmotools}
Performing detailed scientific analysis in post-processing presents a major challenge at exascale, where permanently saving high-resolution particle snapshots at multiple time steps is both prohibitive to store and computationally impractical. A key innovation in \CRKHACC was the development of a complete and fully GPU-accelerated in~situ analysis pipeline. By analyzing data directly on the device during runtime, we eliminate the need to offload and store massive intermediate datasets, while still producing comprehensive scientific outputs.

A central component of this pipeline is the support for clustering-based analysis methods such as DBSCAN~\cite{ester1996density} and friends-of-friends (FOF) halo finding~\cite{davis1985,klypin1983}, as shown in the bottom right panel of Figure~\ref{fig:schematic}. These algorithms determine where halos are located in the simulation, facilitate detection of all galaxies that have formed, and are used to perform mock-survey measurements. 

Cluster finding is computationally intensive, requiring efficient spatial search and neighborhood queries across potentially hundreds of millions of particles per rank. To enable this at scale, we collaborated in the co-development of the publicly available ArborX library~\cite{arborx2020,prokopenko2024}, which provides GPU-native spatial indexing and traversal routines. Combined with the particle overload approach discussed previously, all clustering analysis can be performed locally and efficiently on each node. 

As a result of these efforts, the in~situ analysis phase is not a bottleneck, and its computational cost is subdominant compared to the short-range force solver (see Section~\ref{subsec:TTS}), even for complex multi-species analyses involving dark matter, gas, and stars. This tight coupling of analysis with simulation enables us to extract scientifically rich datasets at full resolution without requiring post-processing of petabytes of raw simulation data. 
\vspace{0.2cm}
\subsubsection{Multi-Tiered I/O}\label{subsec:IO}

Once all computation on the GPU is completed, including short-range force evaluations and on-the-fly analysis, the resulting particle data must be written to the PFS. The majority of I/O involves writing full particle checkpoints ($\sim150-180$\,TB) after each PM step, which is necessary to minimize potential data loss from machine failures. The mean time to interrupt (MTTI) of modern exascale and large-scale commercial AI systems is typically a few hours~\cite{kokolis2024}, so repeated checkpoints are necessary, especially for full machine runs.  

To achieve efficient throughput, we employ a multi-tiered I/O strategy (bottom left panel of Figure~\ref{fig:schematic}). First, each node performs synchronized writes to local NVMe (Non-Volatile Memory Express) solid-state storage, which offers significantly higher bandwidth than the shared parallel file system. Then, a background thread is launched on each node to asynchronously transfer the resulting files to the PFS using low-level operating system move commands.  
Moreover, additional background threads simultaneously remove outdated checkpoints (using a time-window function) as the simulation progresses to avoid storage buildup. 

This approach has several advantages: file contention is avoided since each node writes exclusively to its own local storage before transferring complete files; the simulation continues uninterrupted while data is asynchronously bled and outdated checkpoints are removed; and the method is node-local and fully decentralized, simplifying coordination and improving robustness. On systems without NVMe, the same procedure can be applied node-locally using RAM disk, which we have also successfully deployed on other supercomputing systems such as Aurora.

As shown in Section~\ref{subsec:TTS}, we found this approach to be highly stable, rarely encountering file system stalls, and were able to write 100\,PB of data aggregated to the \Frontier PFS (\emph{\Orion}) with an effective sustained bandwidth of 5.45\,TB/s without directly interfacing with the Lustre PFS during the most latency-sensitive phases of simulation. Given that the theoretical peak bandwidth of \Orion is 4.6\,TB/s~\cite{atchley2023}, our multi-tiered strategy exceeded the bandwidth achievable via direct PFS writes, delivering higher sustained throughput without compromising simulation stability.

\section{Performance Measurement Methodology}\label{sec:perf_measured}

\subsection{HPC Systems and Testbeds}\label{subsec:testbeds}

All simulation and scaling measurements were performed on the OLCF \Frontier supercomputer.\footnote{\url{https://www.olcf.ornl.gov/frontier/}} Each of the 9{,}408 \Frontier nodes consist of a 64-core \AMD EPYC 7A53 “Trento” CPU with 512\,GB of DDR4 memory and four \AMD \emph{Instinct}$^\text{TM}$ \MIX GPUs. The \MIX is composed of two Graphics Compute Dies (GCDs), each capable of delivering 23.9\,\TFLOPS of unpacked FP32 vector instructions connected to 64\,GB of HBM2e memory. Node-local SSD storage includes two NVMe M.2 drives with a combined capacity of $\sim$3.5\,TB, providing sustained read and write bandwidths of 8\,GB/s and 4\,GB/s, respectively. The simulation campaigns used 9{,}000 \Frontier nodes ($\mathord{>}95\%$ of the full system), yielding a theoretical maximum performance of 1.720\,\EFLOPS (FP32) and an aggregate of 36\,TB/s of node-local SSD write bandwidth. Frontier's interconnect is a three-hop Slingshot~11 dragonfly topology connected to the Lustre-based \Orion parallel file system, capable of theoretical peak bandwidths of 5.5\,TB/s (read) and 4.6\,TB/s (write) for large-file workloads~\cite{atchley2023}.

Portability tests on \Intel hardware were carried out on the ALCF \Aurora supercomputer\footnote[2]{\url{https://www.alcf.anl.gov/aurora}}, using nodes with two 52-core Intel Xeon CPU Max Series (codenamed Sapphire Rapids) and six Intel Data Center GPU MAX 1550 (codenamed Ponte Vecchio, PVC) devices. A PVC die consists of two compute tiles, each delivering approximately 22.5\,\TFLOPS of FP32 performance, with access to 64\,GB of HBM2e memory. 

\Nvidia hardware measurements were performed on H100 GPU nodes at the Argonne Joint Laboratory for System Evaluation (JLSE).\footnote[3]{\url{https://www.jlse.anl.gov/}} Each node consists of two 48-core \Intel Xeon Platinum 8468 CPUs and four \Nvidia H100 SXM5 GPUs, each sustaining 66.9\,\TFLOPS of FP32 throughput and paired with 80\,GB of HBM3 memory.

\subsection {FLOPs Measurements}\label{subsec:flops}

FLOP performance measurements on \AMD hardware were obtained using rocprof (ROCm 6.3.1), sampling profile counters for FP32 add, multiply, fused multiply-add (FMA), and transcendental operations.\footnote[4]{FMAs are counted as two operations; transcendental operations are counted as one.} Similarly, \Nvidia measurements were gathered using ncu (CUDA 12.8), and \Intel measurements with GTPin (oneAPI 2025.0.0). Kernel timings on all platforms were extracted using \texttt{MPI\_Wtime}.

Peak FLOP rates were determined by profiling the GPU kernel with the highest measured FP32 operation throughput. For the \CRKHACC solver, this corresponds to the compute kernel responsible for calculating high-order SPH correction coefficients. Sustained FLOP rates were measured by accumulating all FP32 operations across the full solver stack and dividing by the total solver wall-clock time. These measurements include not only the hydrodynamics and gravity force solvers, but also all astrophysical subgrid models, tree-walk operations, interaction list assembly, and memory transfers to and from the device.

We define GPU utilization as $P_\text{measured} / P_\text{hardware}$, the ratio of achieved to theoretical peak FP32 throughput. The hardware-specific FP32 peak rates used in this calculation are listed in Table~\ref{tab:gpuproperties}. An analysis of GPU utilization across architectures and within the full \FRONTIERE simulation is presented in Section~\ref{subsec:perf_port}.

\begin{table}[tb]
\caption{GPU Specifications}
\label{tab:gpuproperties}
\vspace{-0.2cm}
\begin{tabularx}{\columnwidth}{@{}lX@{}}
\toprule
    \textbf{Device} & \textbf{Peak Single Precision} (TFLOPs) \\
\midrule
    AMD MI250X           &  23.9 (per GCD)   \\
    Intel Max 1550 (PVC) &  22.5 (per tile)   \\
    NVIDIA SXM5 H100     &  66.9    \\
\bottomrule
\end{tabularx}
\end{table}

For the full machine run, we assign one GPU tile to its own MPI process and execute 8 MPI processes on each node for a total of 9{,}000 nodes. To obtain performance data, we profile one MPI rank on each node, multiply the obtained performance data by 8, and sum over all nodes in the run. The max time across all ranks is (conservatively) used for system-wide FLOP measurements. Distributions of the performance per rank are shown in Section~\ref{subsec:perf_port}.

\subsection{High and Low Redshift Performance}\label{subsec:redshifts}

\begin{figure}
    \includegraphics[width=\linewidth]{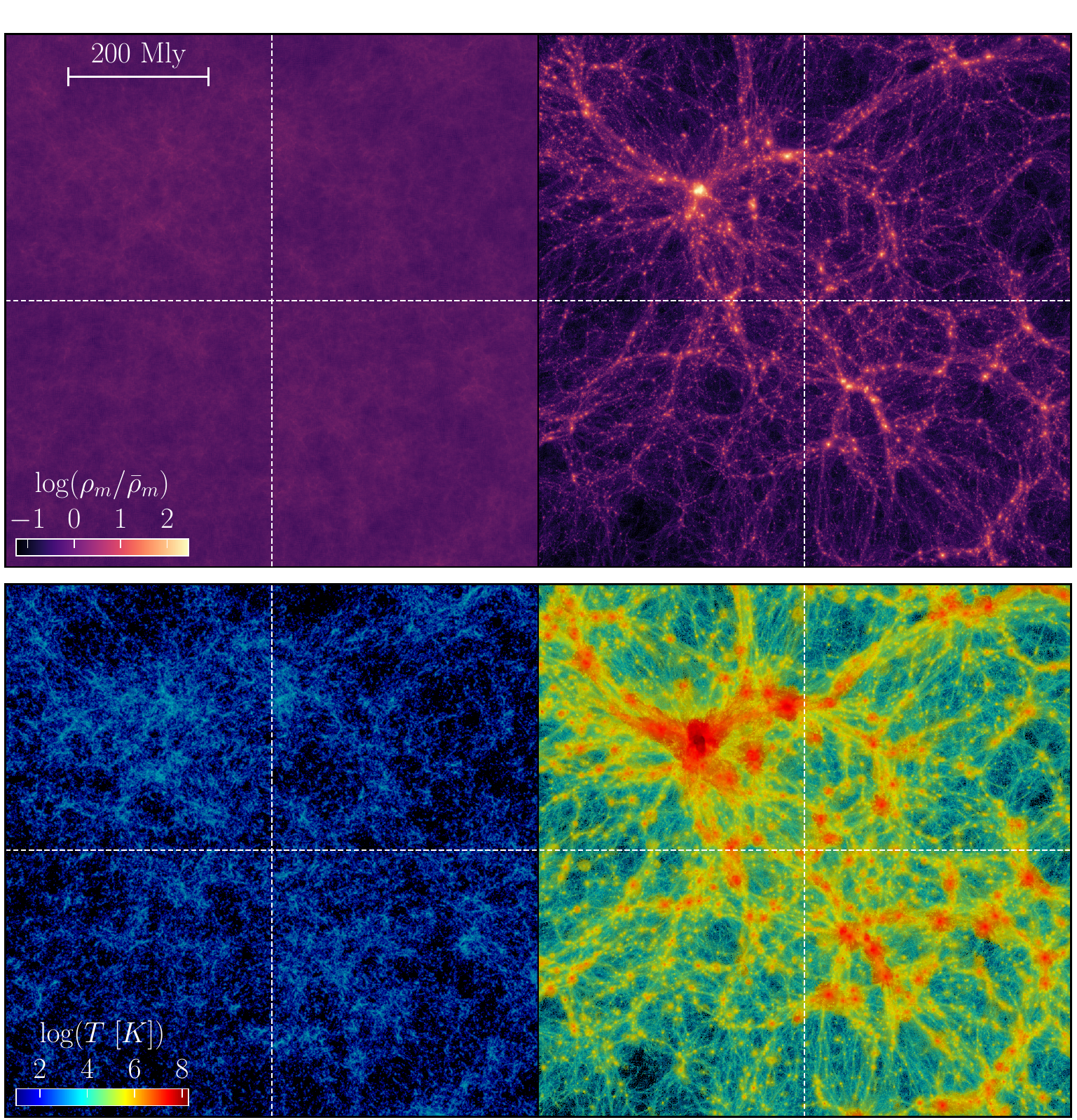}
    \caption{Slices of total matter density (top panels) and gas temperature (bottom panels) from four ranks of the \FRONTIERE simulation at high redshift ($z = 9$; early universe, left) and low redshift ($z = 0$; late universe, right). Dashed lines indicate rank boundaries.}
    \label{fig:highlowz}
\end{figure}

Cosmological simulations present a unique performance challenge due to their need to resolve a large spatial dynamic range throughout the simulation domain over the entire evolution history of the Universe. Accordingly, the nature of the workload evolves significantly over time. As shown in Figure~\ref{fig:highlowz}, the early (high redshift) homogeneous universe is relatively uniform, and computational work is well-balanced across nodes. At late times (low redshift), matter becomes highly clustered, and the computational load becomes increasingly uneven, particularly impacted by stochastic astrophysical feedback models in dense regions that inject significant amounts of energy. To investigate performance across these contrasting regimes, we measured GPU device utilization on 9{,}000 ranks during both early- and late-time simulation phases.

In the high-redshift (high-$z$) phase, we measure both the per-node device utilization and the overall system performance, reporting machine peak (513.1\,PFLOPs) and sustained (420.5\,PFLOPs) rates. 
At low redshift (low-$z$), where strong clustering leads to adaptive time stepping and node-to-node variability in workload, we conducted two performance measurements. First, we measured full-step performance under typical asynchronous integration conditions, capturing the real-world execution profile. Second, to evaluate absolute performance potential in this more complex regime, we conducted a ``low-$z$ Flat'' measurement in which all nodes were artificially synchronized to follow the deepest local timestep. This allowed us to assess GPU efficiency and solver throughput in a controlled but representative late-time workload. The results are summarized in Section~\ref{subsec:perf_port}.  

\subsection{I/O Bandwidths}\label{subsec:IO_band}

\CRKHACC uses a multi-tiered I/O strategy that combines synchronized node-local writes with asynchronous bleeding to the parallel file system, \Orion. Unless otherwise specified, we report performance for the most demanding I/O operation, a full particle checkpoint. Each checkpoint consists of all four trillion particles, including the overloaded ``ghost'' regions, producing approximately 150\,--\,180\,TB of data per output, which is written after every simulation step.

The local storage bandwidth is measured using the total time required to complete the writes on all nodes. For PFS performance, each node records the duration of its asynchronous copy to \Orion, and the effective write bandwidth is computed using the maximum time reported across all nodes. The total output size is calculated by adding the write volume across all ranks over the full course of the simulation.

\section{Performance Results}\label{sec:results}

\subsection{Parallel Scaling and Performance}\label{subsec:scaling}

\begin{figure}
    \includegraphics[width=\linewidth]{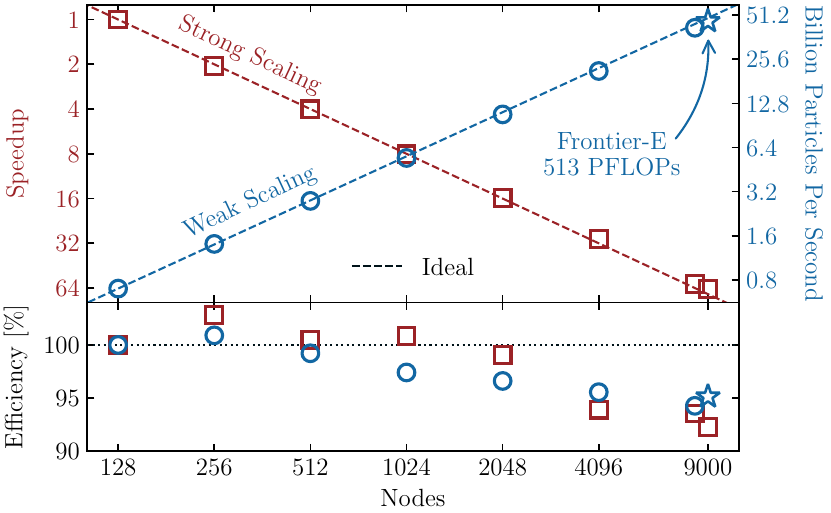}
    \caption{Strong (left axis, in red) and weak (right axis, in blue) scaling from 128 to 9{,}000 nodes on \Frontier, with the lower panel showing efficiency relative to the ideal case. Weak scaling is presented in terms of the number of particles processed per second by the solver. The \FRONTIERE problem size is indicated by the star (46.6 billion particles per second), where we achieved 513.1\,PFLOPs peak and 420.5\,PFLOPs sustained performance.}
    \label{fig:scaling}
\end{figure}
 
Figure~\ref{fig:scaling} shows strong and weak scaling results from 128 to 9{,}000 nodes on \Frontier. For weak scaling, the particle count and volume per rank are held fixed as we scale up to the full \FRONTIERE configuration of $2 \times 12{,}600^3$ particles on 9{,}000 nodes. For strong scaling, the total problem size is fixed at $2 \times 3{,}840^3$ particles, the same size used in the 256-node weak-scaling configuration, while the number of nodes is increased up to 9{,}000. To account for spatial overloading, the results are proportionally adjusted. 

In both cases, we measure the average time spent in the solver (short-range plus spectral components) across four high-redshift steps. We achieve strong and weak scaling efficiencies of 92\% and 95\%, respectively, across nearly two orders of magnitude in node count.

Weak scaling is the most relevant metric for cosmological simulations, where the goal is to grow the problem size in proportion to available computational resources. To emphasize this, we plot the number of particles processed per second rather than time-to-solution, which would remain flat under ideal weak scaling. On 9{,}000 nodes, we process 46.6 billion particles per second -- equivalent to advancing one full high-redshift timestep for all four trillion particles of \FRONTIERE in just a few minutes. The measured peak and sustained full-machine performance are 513.1\,PFLOPs and 420.5\,PFLOPs, respectively.

\subsection{Time-to-Solution and I/O}\label{subsec:TTS}

\begin{figure}
    \includegraphics[width=\linewidth]{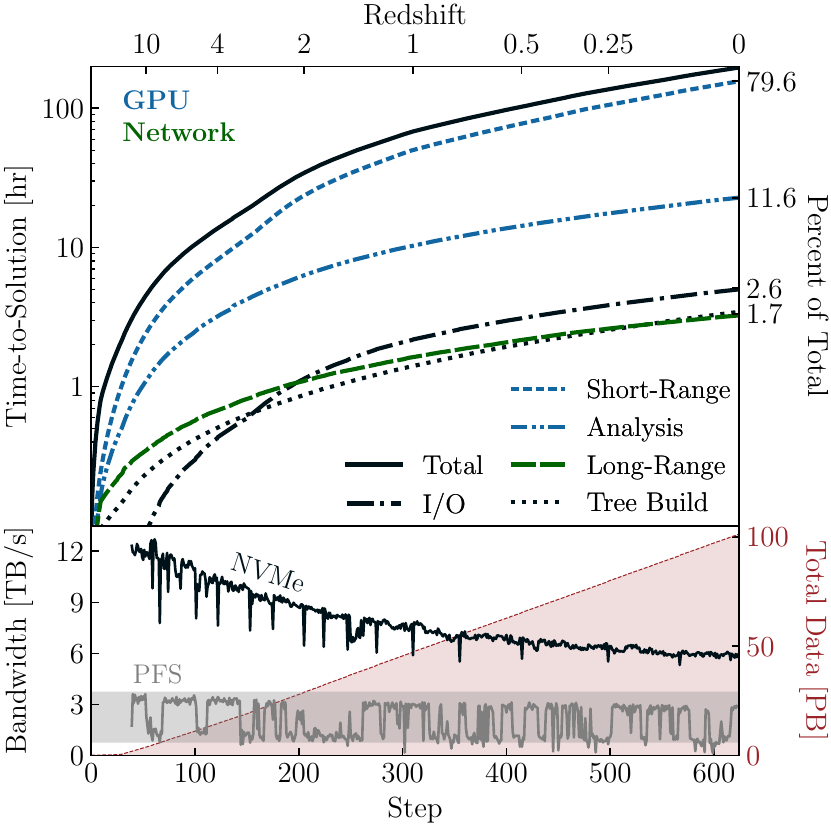}
    \caption{\textbf{Top:} Cumulative time-to-solution of the \FRONTIERE simulation, along with individual timers for the short- and long-range solvers, I/O, tree construction, and analysis [$\sim$2.8\% of the simulation time is in global reductions and miscellaneous software not individually visualized]. Note that redshift decreases non-linearly with cosmic time, so late stages of the simulation span a larger fraction of the Universe’s age. \textbf{Bottom:} NVMe SSD and PFS bandwidth of the multi-tiered I/O strategy, with the gray band bracketing the 0.75\,--\,3.75\,TB/s PFS bandwidth. The shaded red region tracks the total data written during the \FRONTIERE run. }
    \label{fig:timing}
\end{figure}

Figure~\ref{fig:timing} shows the cumulative time-to-solution (TTS) for the \FRONTIERE run over 625 PM timesteps, each of which can include up to thousands of local substeps, spanning the full redshift evolution of the Universe.
Because the relationship between redshift and cosmic time is highly non-linear, a much greater fraction of the Universe’s history is integrated at low-$z$ (toward the end of the simulation), compounding an already more demanding workload, as described in Section~\ref{subsec:redshifts}.

The total wall-clock time was 196 hours, amounting to just over 1.7 million node-hours on \Frontier~-- achieving the target throughput of completing a flagship simulation in approximately one week.  For reference, a gravity-only simulation with an identical configuration completed in just under 12 hours, making the hydrodynamic run approximately 16 times more expensive. These results highlight not only the computational overhead introduced by gas dynamics, but also the capabilities of exascale systems, which can now perform former state-of-the-art gravity-only simulations in half a day.

The detailed timing breakdown in Figure~\ref{fig:timing} reveals several important features of the \CRKHACC execution profile. The short-range force solver dominates the compute cost, contributing 79.6\% of the total time, followed by in~situ analysis at 11.6\%. In total, 91.2\% of the runtime is spent on the GPU, an essential milestone for achieving efficient exascale performance. Given that our goal was to increase the complexity of the problem by more than an order of magnitude, Amdahl’s law dictates that at least 90\% of the workload must be GPU-accelerated to maintain overall efficiency. 

For comparison, assuming similar per-FLOP performance, running the entire simulation on the 3rd Gen Trento CPU cores of \Frontier would result in a wall-clock time of roughly a year! This contrast underscores the importance of GPU acceleration and highlights the advantage of using a fully GPU-optimized code in a domain where such architectures are rarely leveraged.

Continuing to examine the timing profile, the execution time for the tree construction and spectral (long-range) force solver was negligible (a combined $\sim$3\%). The FFTs are performed on global grids of two trillion cells, so minimizing their MPI communication overhead and frequency is critical -- enabled by both a performant FFT distribution implementation and the coarse PM time stepping afforded by the separation-of-scales approach. Additionally, the tree solver is memory-bandwidth bound and is designed to be built only once per PM time step to reduce construction cost. The minimal combined wall-clock time indicates that both architectural design elements are functioning optimally at scale.

I/O accounts for just 2.6\% of the total runtime, a major achievement for writing a total of 100\,PB of data. As highlighted in the lower panel of Figure~\ref{fig:timing}, the multi-tiered I/O strategy was essential in avoiding bottlenecks. High-bandwidth synchronous writes to node-local NVMe drives were followed by asynchronous bleeding to the parallel file system. As the simulation progressed, the data size imbalance across nodes grew to nearly a factor of two, reducing the effective synchronized NVMe write bandwidth by the same factor relative to high-redshift performance. Periodic drops in NVMe bandwidth were primarily due to analysis output steps, where ranks simultaneously read and wrote multiple datasets to local SSDs, temporarily reducing effective write speed by up to 30\%. Even so, node-local write performance remained high, with bandwidths approaching~6\,TB/s toward the end of the run. PFS bandwidth also varied due to complex I/O patterns and Lustre performance variability, but still sustained between 0.75 and 3.7\,~TB/s to \Orion.

Using the 6\,--\,12\,TB/s of node-local SSD bandwidth, we routinely wrote 150\,--\,180\,TB checkpoint files in tens of seconds, while asynchronous background bleeds to \Orion were completed in at most minutes. For fault tolerance, a full checkpoint was written at every timestep. Combined with the complex scientific outputs ($\sim$12\,PB), this resulted in over 100\,PB of data written during the simulation. Dividing the total data volume by the cumulative I/O runtime ($\sim$5.1 hours) yields an effective write bandwidth of 5.45\,TB/s. Given that the theoretical peak write bandwidth of \Orion is 4.6\,TB/s~\cite{atchley2023}, our measured multi-tiered I/O bandwidth exceeds the peak capability of direct-to-PFS writes, demonstrating sustained, high-throughput, fault-tolerant output for a complex scientific pipeline.

In summary, all primary application components are performing optimally at scale -- enabling the necessary throughput to complete the simulation in just over a week. Over 90\% of the total runtime is spent on the GPU, which is critical for high device utilization on exascale systems. The remaining non-compute-bound operations and I/O were optimized to be subdominant, despite performing distributed FFTs on more than two trillion cells and writing more than 100\,PB of data.

\subsection{Utilization and Portability}\label{subsec:perf_port}

\begin{figure}
    \includegraphics[width=\linewidth]{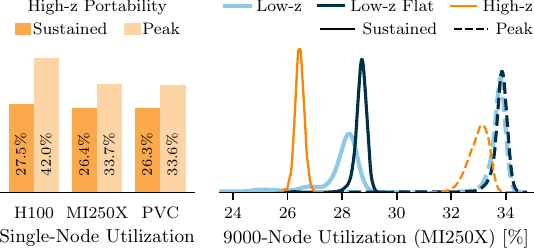}
    \caption{Device utilization measurements. \textbf{Left:} Single node, high-z measurement across three GPU vendors (\Nvidia, \AMD, \Intel).  \textbf{Right:} \FRONTIERE full machine distribution measurements at both high and low redshifts. Low-z Flat measures utilization when all nodes were artificially synchronized to the same time integration depth. }
    \label{fig:utilization}
\end{figure}

Achieving high sustained performance on exascale systems requires extensive GPU acceleration. As discussed in Sections~\ref{subsec:flops} and~\ref{subsec:redshifts}, we quantify the solver performance across hardware architectures and simulation phases by measuring device utilization at both high and low redshifts. Device utilization is defined as the ratio of measured floating-point operations to the theoretical peak FLOP rate of a given GPU.

Figure~\ref{fig:utilization} presents the device utilization of \CRKHACC across different GPU architectures, as well as during the full \FRONTIERE simulation. The left panel shows single-node utilization measurements on NVIDIA H100, Intel PVC, and AMD MI250X GPUs. The sustained utilization achieved by the solver is consistent across all three platforms, with slightly higher peak performance observed on \Nvidia hardware. This demonstrates that the core GPU-resident components of \CRKHACC are GPU-portable and maintain high efficiency across vendor architectures. A detailed performance portability evaluation is provided in Ref.~\cite{rangel2023}.

The right panel of Figure~\ref{fig:utilization} shows device utilization measured for each profiled rank across the full 9{,}000-node \FRONTIERE run at both high and low redshift. At high redshift, where the particle distribution is relatively homogeneous and the workload is well balanced, we observe a peak (high\nobreakdash-$z$) per-GPU utilization of approximately 33\% and a sustained value of 26.5\%, consistent with the single-node runs.
As the simulation progresses to low redshift and the Universe becomes increasingly clustered, the local work per GPU increases, leading to improved per-GPU performance. In this regime, peak (low-$z$) utilization rises to just under 34\%, with sustained utilization reaching 28\%.
However, the distribution of utilization across ranks broadens at low redshift due to variation in workload and timestep depth across the solver.

To isolate the effect of this imbalance, we also ran an artificial ``Flat'' low-redshift configuration in which all ranks were forced to use the same synchronized time step. This removed per-rank time integration variability and produced a much tighter utilization distribution. The similarity in average performance between the Flat and native cases indicates that the timestep adaptivity is functioning as intended and does not introduce significant performance degradation -- even in the most computationally demanding phase of the simulation.

Taken together, these results demonstrate that the solver maintains strong and consistent GPU performance across vastly different dynamical regimes, and demonstrates GPU-portability across hardware vendors. The combination of the adaptive time stepper and the GPU-resident tree solver has proven effective in resolving complex, localized physical processes without compromising efficiency. Even with the significant per-rank time integration required at low redshift, the architecture sustains high device utilization and scalability, underscoring the robustness of the solver design for demanding, physics-rich workloads.

\section{Implications}\label{sec:conclusion}

\FRONTIERE represents the first cosmological hydrodynamic simulation of its kind, achieving survey-scale predictions on par with previous state-of-the-art gravity-only simulations, while incorporating significantly more physical modeling at the trillion-particle scale. Prior large-volume hydrodynamic simulations were at minimum an order of magnitude smaller. The \FRONTIERE simulation provides the resolution, physical realism, and statistical power needed to support next-generation surveys, enabling joint predictions across cosmological observables and full-sky, multi-wavelength modeling.

Achieving this capability required a number of critical innovations as detailed in Section \ref{sec:innovation}. First, a separation-of-scales strategy was employed to decouple long-range and short-range interactions, allowing the latter to remain node-local. Second, roughly fifty short-range kernels (including hydrodynamics, gravity, and astrophysical feedback modules) were fully optimized for GPU execution using customized tree algorithms and the novel warp-splitting approach. Third, all in~situ analyses were executed directly on the GPU to avoid costly transfers and to maintain performance. Fourth, a multi-tiered I/O methodology was developed to leverage node-local SSDs for fast checkpointing and asynchronous bleeding to the parallel file system. 

As shown in Section~\ref{sec:perf_measured}, these innovations enabled near-ideal scaling, with measured peak and sustained FLOP rates of 513 and 420\,PFLOPs processing over 46 billion particles per second. The simulation was completed in just over a week of wall-clock time, with consistently high GPU utilization across very different computational regimes (i.e., high vs. low redshift). In total, over 100\,PB of data were written in negligible runtime, supported by highly efficient and fault-tolerant I/O infrastructure.

\FRONTIERE establishes a new baseline for what is achievable in cosmological simulation, paving the way for even more ambitious efforts to follow. With the capabilities now demonstrated on exascale systems, future runs can pursue higher resolution, improved physical models, and more targeted predictions for specific observational goals. The achieved throughput not only enables flagship simulations, but also advances the scale and fidelity of ensemble campaigns -- important for building emulators, incorporating AI/ML approaches, calibrating models, and estimating covariances -- where greater statistical power directly translates into improved scientific constraints.

The computational strategies developed here have broad relevance beyond cosmology. The short-range force optimizations, I/O methods, and modular solver design are readily generalizable to other particle-based domains such as plasma physics, molecular dynamics, and astrophysical fluid modeling. As future machines continue to increase in GPU density while exhibiting shorter mean times between failures, the resilience and portability demonstrated by \CRKHACC will become increasingly important. Efficient, high-frequency checkpointing, enabled by hierarchical I/O, offers one viable path to ensuring fault tolerance on future large-scale systems, and stresses the importance of node-local persistent storage. 

\FRONTIERE marks the beginning of a new generation of simulations that can fully exploit current and emerging hardware capabilities to address the most profound challenges in cosmology and large-scale structure formation.

\section*{Acknowledgment}

The authors thank Nicholas Malaya, Noah Wolfe, Brian Cornille, Karl W. Schulz, and the AMD performance and application teams, as well as John Pennycook, Zhiqiang Ma, Varsha Madananth, and the Intel performance team. We acknowledge the staff at the Oak Ridge and Argonne Leadership Computing Facilities and at NERSC for their support. This research was supported by the Exascale Computing Project (17-SC-20-SC), a collaborative effort of the U.S. DOE Office of Science and NNSA and by the U.S. Department of Energy, Office of Science, Office of Advanced Scientific Computing Research and Office of High Energy Physics, Scientific Discovery through Advanced Computing (SciDAC) program. This research used resources of the Oak Ridge Leadership Computing Facility at the Oak Ridge National Laboratory, which is supported by the Office of Science of the U.S. Department of Energy under Contract No. DE-AC05-00OR22725; the Argonne Leadership Computing Facility (Contract No. DE-AC02-06CH11357); and the National Energy Research Scientific Computing Center (Contract No. DE-AC02-05CH11231). CAFG is supported by NSF (AST-2108230, AST-2307327), NASA (21-ATP21-0036, 23-ATP23-0008), and STScI (JWST-AR-03252.001-A). Lastly, NF thanks his mother for help in improving the clarity and readability of the paper.

\bibliography{bib}

\end{document}